\documentstyle[aps,twocolumn,epsfig]{revtex}

\newcommand{\beq}{\begin{equation}}

\newcommand{\eeq}[1]{\label{#1} \end{equation}}

\newcommand{\beqar}{\begin{eqnarray}}

\newcommand{\eeqar}[1]{\label{#1} \end{eqnarray}}

\newcommand{\bmath}{\begin{displaymath}}

\newcommand{\emath}{\end{displaymath}}

\newcommand{\bitem}{\begin{itemize}}

\newcommand{\eitem}{\end{itemize}}

\begin{document}






\wideabs{
\title{\Large \bf Strong Color Field Baryonic Remnants in
Nucleus-Nucleus Collisions at 200A GeV.} 
\author{~V.~Topor~Pop$^1$,~M.~Gyulassy$^2$, 
~J.~Barrette$^{1}$, and~C.~Gale$^{1}$}

\address{
$^1$McGill, University, Montreal, Canada, H3A 2T8\\
$^2$Physics Department,
Columbia University, New York, N.Y. 10027\\
}

\date{September 26, 2005}

\maketitle


\begin{abstract}
The effects of strong color 
electric fields (SCF) on the baryon production  at RHIC
are studied in the framework of HIJING/B\=B (v2.0) model.
The particle species dependence of nuclear 
modification factors (NMF) are analyzed for Au+Au collisions at 200A GeV.
A doubling of the string tension leading to a modification 
of the strangeness suppression according to Schwinger mechanism
is shown to provide an alternate
explanation to coalescence models for the interpretation 
of the observed baryon and meson production 
at moderate $p_T$ and results in a predicted
enhancement in the (multi)strange (anti)hyperon production.\\

\end{abstract}

}



\section{Introduction}

While the phase transition from hadronic degree of freedom to
partonic degrees of freedom (quarks and
gluons) in ultra-relativistic nuclear collisions is a central focus of
recent experiments at  the Relativistic Heavy Ion Collider (RHIC),
data on baryon and hyperon production
has revealed interesting and unexpected 
features at RHIC that may be of novel dynamical origin instead.
The so called {\em baryon/meson anomaly} 
\cite{Adcox:2001mf,Adler:2003kg,Vitev:2001zn,Gyulassy:2003mc} 
is observed as a large enhancement of the baryon to meson ratio
and a large difference of the nuclear modification
factor (NMF) between total charged \cite{Adcox:2001jp}  
and neutral pions ($\pi^0$) \cite{David:2001gk} 
at moderate  transverse momenta ($2<p_T<5$ GeV/c).

In a previous paper \cite{prc70_top04} we studied 
the possible role of topological baryon junctions\cite{Kharzeev:1996sq} 
in  nucleus-nucleus collisions.
We have shown in the framework of HIJING/B\=B v2.0 model,
that junction-antijunction (J\=J) loops 
with an enhanced {\em intrinsic
transverse momentum}  $k_T \approx $ 1 GeV/c,
a default string tension $\kappa_{0}$ = 1 GeV/fm,
and a di-quark suppression factor (PARJ(1)=$\gamma_{qq}$=0.07)
provide a partial explanation of the baryon/meson 
anomaly \cite{prc70_top04}.
That model therefore provides an alternative dynamical 
explanation of the data to recombination models \cite{muller03}.  
Within  HIJING/B\=B v2.0 \cite{prc70_top04} 
one of the main assumptions is: 
the strings could survive and fragment \cite{miklos_zf_91,biro_91}, 
and in particular populate the mid to
low $p_T$ range.
In contrast, in the recombination picture \cite{muller03} or 
in hydrodynamical approach \cite{heinz_03}
all coherent strings are assumed to become rapidly incoherent
resulting in rapid thermalization.

In this paper we explore further dynamical effects associated with
long range coherent fields (i.e strong color fields, SCF)
including baryon junctions \cite{Kharzeev:1996sq} and loops 
\cite{Vance:1999pr}
that may arise in nuclear reactions. Our emphasis here
will be on the novel baryon observables measured at RHIC.
In nucleus-nucleus collisions the color 
charge excitations may be considerably
greater than in nucleon-nucleon collisions due to the almost
simultaneous interaction of several participating nucleons in a row
\cite{Biro:1984cf,miklos_zf_91}. 
Molecular dynamics models \cite{ropes_sorge,bleicher_99,bass_00}
have been used to study the effects of color ropes
as an  effective description of the
non-perturbative, soft gluonic part of QCD
\cite{soff_prl03,soff_jpg04,soff_jpg04_l35}. Strangeness enhancement
\cite{rafelski_82,rene_04,caines05,cgreiner_02,armesto96,bleicher_96,antai97,soff_jpg01,brat_prc04}, 
strong baryon transport 
\cite{csernai01}, and increase of intrinsic $k_T$ \cite{soff_jpg04} 
are all expected consequences of SCF.
This can be  modeled  in microscopic models as
an increase of the effective  string tension that controls the
$q\bar{q}$ and $qq\overline{qq}$ pair creation rates
and strangeness suppression factors \cite{Biro:1984cf}.

For a uniform chromoelectric flux tube with field (E) the probability 
to create a pair of quarks with mass (m), effective charge (e),
and transverse momentum ($p_T$) per unit time per unit volume 
is given by \cite{nussinov_80} :
\begin{equation}
P(p_T)\,d^2p_T 
=-\frac{|eE|}{4 \pi^3} ln \Bigg\{ 1-exp\left[-\frac{\pi(m^2+p_T^2)}
{|eE|} \right] \Bigg\} \,\,d^2p_T
\end{equation}
The integrated probability ($P_m$) reproduces the classical Schwinger
results \cite{schwinger_51} , derived in spinor quantum
electrodynamics (QED) for $e^{+}e^{-}$ production rate,
when the leading term in Eq. 2 is taken into account, i. e.:
\begin{equation}
P_m=\frac{(eE)^2}{4 \pi^3} \sum_{n=1}^{\infty}
\frac{1}{n^2}exp\left(-\frac{\pi\,m^2n}{|eE|}\right)
\end{equation}
In general in microscopic string models 
the heavier flavors (and di-quark) are suppressed 
according to Schwinger formula 
\cite{schwinger_51}:
\begin{equation}
\gamma_{Q}=\frac{P(Q\bar{Q})}{P(q\bar{q})}=
exp\left(-\frac{\pi(m_{Q}^2-m_q^2)}{\kappa}\right)
\end{equation}
where $\kappa=|eE|$ is the {\em string tension};
$m_{Q}$ is the effective
 quark mass; (Q=s for strange quark; Q=qq for a di-quark),
and q=u,d are the light nonstrange quarks.
 
In the case of quark-gluon plasma (QGP) creation 
it is necessary to modify 
the dynamics of particle vacuum production at short time scales,
and the abundance of newly produced particle may deviate considerably 
from the values obtained for the constant field \cite{zabrodin_04}. 
Two possible processes may lead to an increase of 
strangeness production 
within the framework of Schwinger mechanism are:
i) increasing the field strength by a modified string tension ($\kappa$)
\cite{Biro:1984cf,soff_jpg04,armesto96,bleicher_96,soff_99},
or ii) dropping the quark masses  
due to chiral symmetry restoration 
\cite{brown_91,soff_99,bleicher_01,bielich_04}.
 A specific chiral symmetry restoration  could be induced by
a rapid deceleration of the colliding nuclei \cite{dima_k05}.
  
Present estimates \cite{pdb_04} of 
the {\em current quark masses} range from:
$m_{u}$ =1.5-5 MeV; $m_{d}$=3-9 MeV, $m_{s}$=80-190 MeV. For
di-quark we consider $m_{qq}$=450 MeV \cite{ripka_04_1}.
Taking  for constituent quark masses
of light non-strange quark $M_{u,d}$= 230 MeV,
strange quark $M_{s}$=350 MeV \cite{amelin_01},  
and di-quark mass $M_{qq}$=550 $\pm$ 50 MeV as in Ref. \cite{ripka_04_1},
it is obvious that the masses of (di)quark and strange quark
will be substantially reduced at the chiral phase transition.
If the QGP is a chirally restored phase of strongly 
interacting matter, 
in this picture the production of strange hadrons will be enhanced 
\cite{bielich_04}.
In this case a possible decrease of the strange quark 
mass would lead to a similar enhancement of the suppression factors,
obtained (in microscopic models) by an increase  of string tension 
\cite{soff_prl03,soff_jpg04,soff_99,zabrodin_04,bleicher_01}.
Moreover, if we consider that Schwinger tunneling could explain
the thermal character of hadron spectra  
and that, due to SCF effects the string tension value $\kappa$ fluctuates
we can define an apparent 
temperature $T=\sqrt{<\kappa>/2\pi}$ \cite{flork_04}.

\section{Outline of the HIJING/B\=B v2.0 model}

Our analysis are performed in the framework 
of HIJING/B\=B v2.0 model \cite{prc70_top04}
that is based on HIJING/B\=B v1.10 \cite{svance99}.
Multiple hard and soft interactions 
proceed as in HIJING v1.37 \cite{wang_97}.
In HIJING/B\=B v2.0 we introduced \cite{prc70_top04}
the possible topology with two junctions \cite{ripka_03}, 
and a new algorithm where  
$J\bar{J}$ loops are modeled by an enhancing di-quark  $p_T$ kick
characterized by a gaussian  width  
of $\sigma_{qq}\,'=f \cdot \sigma_{qq}$ = 1.08 GeV/c, 
with $\sigma_{qq}$=0.360 GeV/c (consistent with PYTHIA 
\cite{pyt_94} default value), i.e. $f$=3, which we fit to best 
reproduce the observed $p_T$ spectrum of the baryons.

Following the equations above, we 
take into account SCF in our model by an 
{\em in medium effective string tension} 
$\kappa > \kappa_0$, which lead to new values for the suppression factors, 
as well as the new effective intrinsic transverse momentum $k_T$ 
\cite{soff_prl03,soff_jpg04,soff_jpg04_l35}. 
This includes: 
i) the ratio of production rates of  
di-quark to quark pairs (di-quark suppression factor),  
$\gamma_{qq} = P(qq\overline{qq})/P(q\bar{q})$,
ii) the ratio of production rates of strange 
to nonstrange quark pairs (strangeness suppression factor), 
$\gamma_{s}=P(s\bar{s})/P(q\bar{q})$,
iii) the extra suppression associated with a diquark containing a
strange quark compared to
the normal suppression of strange quark ($\gamma_s$),
$\gamma_{us}=(P(u\bar{u}s\bar{s})/P(u\bar{u}d\bar{d}))/(\gamma_s)$, 
iv) the suppression of spin 1 diquarks relative to spin 0 ones
(appart from the factor of 3 enhancement of the former based on
counting the number of spin states), $\gamma_{10}$, and 
v) the (anti)quark ($\sigma_{q}''=\sqrt{\kappa/\kappa_0} \cdot \sigma_{q}$)
and  (anti)di-quark ($\sigma_{qq}''= \sqrt{\kappa/\kappa_0} \cdot f
\cdot \sigma_{qq}$) gaussian  width.
These parameters correspond to $\gamma_{qq}$=PARJ(1), 
$\gamma_{s}$=PARJ(2), $\gamma_{us}$=PARJ(3), $\gamma_{10}$=PARJ(4),
and $\sigma_{qq}$= PARJ(21) of the JETSET7.3 subroutines \cite{pyt_94}.
Our calculations are based on the assumption 
that the effective enhanced string tension ($\kappa$), 
in both basic ropes ($q^n-\bar{q}^n$) and junction ropes
($q^n-{q}^n-q^n$) are the same. 
For elementary $n$ strings and junctions this ansatz is supported by 
baryon studies \cite{takahashi_05}.  
A different approach to baryon production without baryon junctions
has been proposed in \cite{armesto96} where 
SCF from string fusion process can lead to 
$(qq)_6-(\bar{q}\bar{q})_{\bar{6}}$ with about double the string tension. 
Both types of SCF configurations may arise but predict different
rapidity dependence of the valence baryons. 
We consider in this version 2.0 of HIJING B$\bar{B}$ 
only  baryon junction rope loops.

There is a debate  in the study of qqq system on the shape 
of $\Delta$-like geometry and $Y$-like geometry 
\cite{ripka_03}, \cite{hoft_04},
and on the stability of  
these configurations for the color electric fields \cite{cgreiner_04_y}.
In both topologies we expect a higher string tension than in an
ordinary $q\bar{q}$ string ($\kappa_Y=\sqrt{3}\,\kappa_0$ and 
$\kappa_{\Delta}=(3/2)\,\kappa_0$).
It was shown  \cite{cgreiner_04_y} that the total string tension 
has neither the $Y$ nor the $\Delta$-like value,
but lies rather in-between the two pictures.
However, the Y configuration appears to be
a better representation of the baryons . 
If two of these quarks stay close together, they behave as a di-quark
\cite{hoft_04}.
In a dual superconductor models of color confinement 
for the Y-geometry the flux tubes converge first toward the centre
of the triangle and there is also another component which 
run in opposite direction. They attract each other and this 
lower the energy of Y-configuration \cite{ripka_03}.

Phenomenological applications are currently based on
Regge trajectory which gives the appropiate relationship 
between the mass M of the hadrons and its spin $J_s$: 
$J_s=\alpha+\alpha_{s}\,\,M^2$,
where $\alpha \simeq 0.5$ is the Regge intercept; 
$\alpha_{s}$ is the Regge slope.
The value of the Regge slope for baryons   
is  $ \alpha_{s}\,\simeq 1\,\,GeV^{-2}$ \cite{svance99} that 
yields a string tension 
( related to the Regge slope, $\kappa_0=1/2 \pi \alpha_{s}$ \cite{wong_hi94})
$\kappa_0 \approx $1 GeV/fm.
This value is taken in our calculations within  HIJING/B\=B v2.0
toghether with the following values for the suppression factors
correponding to the constituent quark masses given above:
$\gamma_{qq}$ =0.02; $\gamma_{s}$=0.30; $\gamma_{us}$=0.40;
$\gamma_{10}$=0.05.
A  broadening in $k_T$ is chosen as: $\sigma_{q}$=0.360 GeV/c and
 $\sigma_{qq}'$ = 1.08 GeV/c (i.e f=3).
HIJING/B\=B v2.0 predictions using this set of parameters
are labeled here by w/o SCF, i.e. without strong 
color field, or by $\kappa_0$=1 GeV/fm.

The multi-gluon exchange processes dominated by Pomeron
exchange in high energy nucleus-nucleus collisions could be described
by a Regge trajectory with a smaller slope  
$\alpha_{s}\,'\, \approx $ 0.45 GeV$^{-2}$ \cite{collins_77}, 
leading to an increase of   
string tension to $\kappa \approx \, 2\,\kappa_0$ \cite{soff_jpg04}, 
corresponding to an increasing values for the suppression factors:
$\gamma_{qq}'$=0.12; 
$\gamma_{s}\,'$=0.55; $\gamma_{us}\,'$=0.63, $\gamma_{10}\,'$=0.12 
as well as of a broadening in $k_T$ , 
$\sigma_{q}''$=0.500 GeV/c and $\sigma_{qq}''$ = 1.5 GeV/c.
The results obtained with this  
set of parameters are labeled here w/SCF, i.e. 
with  strong color field, or by $\kappa$= 2 GeV/fm.
We note, that the increase of ``intrinsic 
$k_T$'' (the gaussian width $\sigma_{qq}''$),
is strongly supported by recent  experimental values reported 
by PHENIX  \cite{rak_04}, which show an increase 
from p+p ($k_{Ty}'$=1.08 $\pm$ 0.05 GeV/c) to  d+Au collisions 
($k_{Ty}''$=1.36 $\pm$ 0.07 $\pm$ 0.12 GeV/c)  
at $\sqrt{s_{NN}}$=200 GeV.

\section{Nuclear Modification Factors}

Here we will concentrate our analysis on species dependence 
of the nuclear modification factors (NMF) $R_{AA}$ 
and $R_{cp}$ in Au+Au collisions at $\sqrt{s_{NN}}$=200 GeV.
$R_{AA}$ is  the ratio of the heavy-ion yield to 
the pp cross section normalized by the number of binary collisions,
while $R_{cp}$ is the ratio of scaled central to peripheral 
particle yield and are defined as in Ref.\cite{prc70_top04}.
For p+p interactions we have used in our calculations, the set of 
optimized parameters from PYTHIA \cite{pyt_94} or HIJING
\cite{wang_97}.

\begin{figure} [t!]
\centering
\epsfig{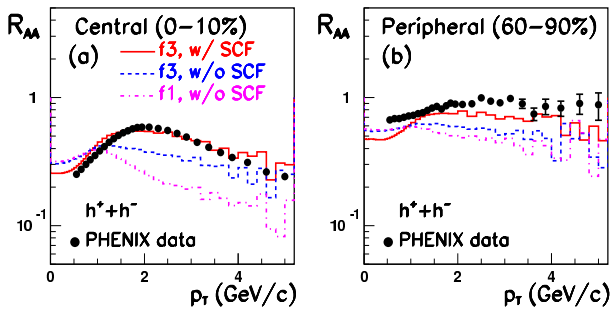} 
\vskip 0.5cm
\caption[Raa vs pt AA] {\small 
(Color online) Comparison of HIJING/B\=B v2.0 predictions 
for $R_{AA}$ of total inclusive charged hadrons,
in central (0-10\% -left) and peripheral (60-90\%-right)
Au+Au collisions at $\sqrt{s_{NN}}$=200 GeV. The results are with  
(solid histograms) and without SCF (dashed histograms).   
The label f3 stands for model calculations assuming $f$=3.
Dashdotted histograms are the results 
without SCF and $f$=1 (label f1).
The data are from PHENIX \cite{Adler:2003au}.
Only statistical error bars are shown.
\label{fig:fig1}
}
\end{figure}

Figure~\ref{fig:fig1}  shows the predicted NMF
$R_{AA}$ for the sum of hadrons for two centralities. 
The results are compared to the data obtained 
by the PHENIX collaboration \cite{Adler:2003au}.
The data at both centralities could not be described  
by assuming only  
a broadening of the {\em intrinsic $k_T$} from 
its standard value (dotted histograms) to 
$\sigma_{qq}'$=1.08 GeV/c (i.e $f$=3, dashed histograms).
The introduction of SCF has an 
effect on the predicted nuclear modification factors
of the total inclusive hadrons and  
results in a better agreement with data (solid histograms).
The data indicate at most a small variation  
with centrality of the factor ``$f$''  consistent with 
the broadening originating at the parton level.

\begin{figure} [hbt!]
\centering
\epsfig{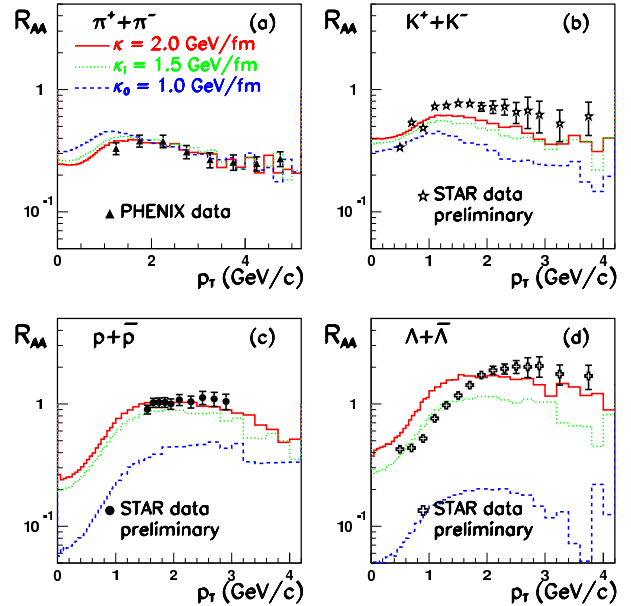} 
\vskip 0.5cm
\caption[R_AA pions,K,p,L0 ] {\small (Color online)
HIJING/B\=B v2.0 predictions for species dependence of NMF ($R_{AA}$)
in central  (0-10\%) Au+Au collisions at $\sqrt{s_{NN}}$=200 GeV:
(a) for charged pions, (b) kaons, (c) inclusive p+\=p, 
(d) inclusive $\Lambda+\bar{\Lambda}$. 
The results are with  
(solid histograms) and without SCF (dashed histograms).   
The dotted histograms are the predictions 
assuming $\kappa_i$=1.5 GeV/fm.
The data are from 
PHENIX \cite{Adler:2003qi} and STAR \cite{rene_sqm04,mironov_sqm04} 
collaborations. Only statistical error bars are shown.
\label{fig:fig2}
}
\end{figure}

In order to better quantify possible effects of strong color 
field on particle production 
we investigate species dependence of NMFs $R_{AA}(p_T)$
for central collisons (Fig.~\ref{fig:fig2}a-d.), 
where higher sensitivity to SCF is expected.
Because of their dominance, the production of pions 
is only moderately modified when we 
consider an increase of the string tension value, 
since the total energy is conserved.
Taking into account SCF effects (solid histograms)
results in changes at moderate $p_T$ 
of less than  $\approx 20\%$ for the pion yield
(Fig.~\ref{fig:fig2}a).
The scaling behaviour in $R_{AA}(p_T)$ of the pions is different 
from those of the sum of protons and anti-protons
(Fig.~\ref{fig:fig2}c).
The pions yield in central events 
is strongly suppressed compared to binary collisions scaling 
($R_{AA}(p_T)$=1). 

The hadron production in  HIJING/B\=B v2.0 is mainly from the  
fragmentation of energetic partons. Thus, 
the observed suppression of pions in central collisions may be
a signature of the energy loss of partons 
during their propagation through the hot and dense matter (possibly QGP)
created in the collisions, i.e {\em jet quenching}.
On the contrary, strange particles are highly sensitive 
to the presence of SCF. 
Kaons (Fig.~\ref{fig:fig2}b) and
lambda (Fig.~\ref{fig:fig2}d), show an increase 
by a factor of two and ten, respectively. 
Such an increase results in a predicted enhancement 
of the lambda yield relative to scaled binary collisions 
as opposed to the strong suppression predicted (and observed) for
pions. These results are consistent with the 
PHENIX data \cite{Adler:2003qi} and preliminary STAR data 
\cite{rene_sqm04,mironov_sqm04}.
In order to study the sensitivity to string tension values 
we also present the results corresponding to an intermediate 
value of the string tension i.e. $\kappa_i$= 1.5 GeV/fm (dotted histograms). 
The discrepancy seen at low $p_T<1.2\, $ GeV/c (solid histograms)
comes from a sizeable contribution from radial flow,
not included in our model. The low $p_T$ region
at RHIC seems to be 
better described within hydodynamical approach \cite{heinz_03}.

\begin{figure} [hbt!]

\centering

\epsfig{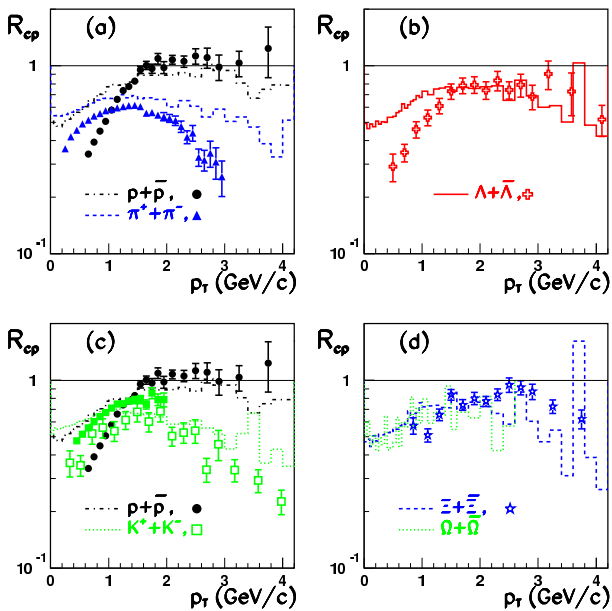} 

\vskip 0.5cm

\caption[R_cp] {\small (Color online)
HIJING/B\=B v2.0 w/SCF ($\kappa=2$ GeV/fm) predictions for
species dependence of NMFs $R_{cp}(p_T)$ 
in Au+Au collisions at $\sqrt{s_{NN}}$=200 GeV.
The results are for scaled (0-10\%)/(60-90\%).
The data are from PHENIX (filled symbols) and from STAR (open symbols).
PHENIX data are scaled (0-10\%)/(60-90\%) \cite{Adler:2003kg}.
STAR data are at slightly different centralities: 
scaled (0-5\%)/(60-80\%) for kaons and 
$\Lambda's$ \cite{Adams:2003am}, and scaled (0-5\%)/(40-60\%) for 
$\Xi$ (preliminary data) \cite{rene_sqm04}.
Only statistical error bars are shown. 
\label{fig:fig3}
}
\end{figure}

\begin{figure} [hbt!]

\centering

\epsfig{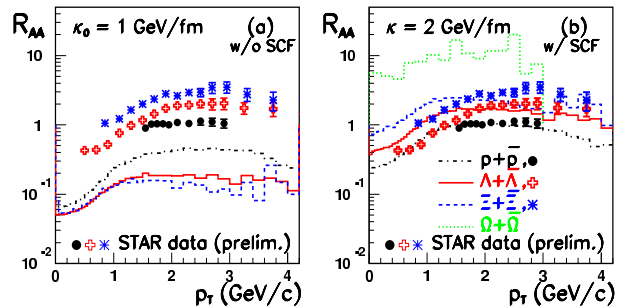} 

\vskip 0.5cm

\caption[R_AA ] {\small (Color online)
HIJING/B\=B v2.0 predictions with SCF (right) 
and without SCF (left) for species dependence of $R_{AA}(p_T)$ 
in central (0-5 \%) Au+Au collisions at $\sqrt{s_{NN}}$=200 GeV.
The preliminary STAR data are from \cite{rene_sqm04,mironov_sqm04}.
Only statistical error bars are shown.
\label{fig:fig4}
}
\end{figure}

The predicted particle dependence is due to the interplay 
between nuclear effects such as jet quenching and shadowing 
and fluctuations of the chromoelectric field 
in the early phase of the reaction. 
Insight into the source of the particle dependence is obtained from 
Fig.~\ref{fig:fig3} and Fig.~\ref{fig:fig4}.
The comparison of the NMF $R_{cp}(p_T)$ for protons and pions
(Fig.~\ref{fig:fig3}a) as well as for kaons
(Fig.~\ref{fig:fig3}c) 
presents a behavior similar to $R_{AA}(p_T)$, i.e shows 
a meson/baryon anomaly that is well described and was  
interpreted in \cite{prc70_top04} as due to a possible 
exotic gluonic mechanism (J\=J loops). 

In Ref.~\cite{muller03} it is suggested that 
the  behavior of $R_{cp}(p_T)$ may 
be interpreted as due to the competition between 
recombination and parton fragmentation.  
The results obtained for $R_{cp}(p_T)$ for the 
strange (Fig.~\ref{fig:fig3}b) and multi-strange particles 
(Fig.~\ref{fig:fig3}d) show 
a small suppression relative to binary scaling,
consistent with the experimental results for $\Lambda$  
\cite{Adler:2003kg,Adams:2003am}, and preliminary data 
for $\Xi$ \cite{rene_sqm04}.

In contrast, the equivalent predictions of $R_{AA}(p_T)$ for 
 protons and (multi)strange particles (Fig.~\ref{fig:fig4}) 
show a predicted strong 
 enhancement of NMF due to SCF effects dependent on 
the mass and strangeness content of the produced particles.

In particular, the model predicts a dramatic 
enhancement in the multi-strange 
(anti-)hyperon production at moderate $p_T$, up to a factor 
of roughly 2 (relative to binary scaling) 
for $\Xi$s and up to a factor of more than 10 for $\Omega$s 
(Fig.~\ref{fig:fig4}b). 
The striking difference between $R_{AA}$ and $R_{cp}$ could be
explained in our model as a consequence of  
SCF which manifest in both central and peripheral
collisions. This shows that there is a clear difference
between using  peripheral Au+Au yields (as in $R_{cp}$), 
or p+p yields (as in $R_{AA}$) as base-line for comparison
with binary collisions scaling ($R_{AA}$=$R_{cp}$=1).

\section{Summary and Conclusions}

In summary, we studied the influence of possible strong 
longitudinal color fields
in particle production in heavy-ion collisions. We modeled SCF effects
within HIJING/B\=B v2.0, by varying the effective  
string tension that controls the
$q\bar{q}$ and $qq\overline{qq}$ pair creation rates
and strangeness suppression factors.
We show that junction-(anti)junction loops and    
a higher string tension $\kappa=2\,\kappa_{0}$
($\kappa_{0} \approx$ 1 GeV/fm), could be important dynamical
mechanisms in the solution of the observed {\em baryon/meson anomaly}.
Our approach has the advantage of correlating many observables
in the same dynamical model including pions, kaon, and baryons 
productions and of predicting 
species dependence of the nuclear modification factors 
$R_{AA}(p_T)$ and $R_{cp}(p_T)$, and the transverse momenta
over a large $p_T$ range.
A greater sensitivity to SCF effects is predicted 
for the nuclear modification factors of (multi)strange hyperons.
In particular, the measurement of $\Omega$ and $\bar\Omega$ yields would
provide an important test of the consistency
of SCF and baryon junction mechanisms at RHIC.

\section{Acknowledgments}

We thank N. Xu and R. Bellwied for helpful discussions
and suggestions throughout this project.

This work was partly supported by the Natural Science and 
Engineering Research Council of Canada and the 
Fonds Nature et Technologies of Quebec.  
This work was supported also by the Director,
Office of Energy Research, Office of High Energy and Nuclear Physics,
Division of Nuclear Physics, and by the Office of Basic Energy
Science, Division of Nuclear Science, of the U. S. Department 
of Energy under Contract No. DE-AC03-76SF00098 and
DE-FG-02-93ER-40764.


\end{document}